\documentclass[aps,pra,twocolumn]{revtex4-1}
\usepackage{amsfonts,amssymb,amsmath}            
\usepackage{lmodern}
\usepackage[]{graphics,graphicx}            
\usepackage{amsthm}
\usepackage{bbold,enumerate,mathtools}
    \usepackage{hyperref}

\newcommand{\ket}[1]{\left | #1 \right\rangle}
\newcommand{\bra}[1]{\left \langle #1 \right |}
\newcommand{\half}{\mbox{$\textstyle \frac{1}{2}$}}

\newcommand{\Tr}{\text{Tr}}

\newcommand{\proj}[1]{\ket{#1}\bra{#1}}
\newcommand{\identity}{\mathbb{1}}
\renewcommand{\epsilon}{\varepsilon}
\begin{document}

\title{Quantum Error Correction for State Transfer in Noisy Spin Chains}
\date{\today}

\author{Alastair \surname{Kay}}
\affiliation{Royal Holloway University of London, Egham, Surrey, TW20 0EX, UK}
\email{alastair.kay@rhul.ac.uk}
\begin{abstract}
Can robustness against experimental imperfections and noise be embedded into a quantum simulation? In this paper, we report on a special case in which this is possible. A spin chain can be engineered such that, in the absence of imperfections and noise, an unknown quantum state is transported from one end of the chain to the other, due only to the intrinsic dynamics of the system. We show that an encoding into a standard error correcting code (a Calderbank-Shor-Steane code) can be embedded into this simulation task such that a modified error correction procedure on read-out can recover from sufficiently low rates of noise during transport.
\end{abstract}

\maketitle

\section{Introduction}

One of the most promising, and furthest progressed, uses of a quantum computer is the quantum simulator, wherein the Hamiltonian of one well controlled and understood quantum system can reproduce the Hamiltonian of another system that we wish to study \cite{Somaroo1999,Friedenauer2008,Gerritsma2010,kim2010,Islam2011}. The challenge for these simulators is the tolerance of noise. In principle, this can be done -- a universal quantum computer can implement gates consisting of the Trotterised Hamiltonian within a fault-tolerant architecture \cite{lanyon2011,jones2012}. However, this digital method of simulation means many of the benefits of the original (analog) Hamiltonian simulation are lost; certainly we can no longer use the comparatively easy route of natural Hamiltonian dynamics. We seek a middle ground; a method for embedding robustness into an analog Hamiltonian simulation. Such a task appears extremely challenging -- even if the system's error channels are as simple as possible (e.g.\ acting locally and independently on each spin), then by the end of the simulation of Hamiltonian $H$, a time $t$ later, the error operator $\hat O$ has propagated to $e^{-iHt}\hat Oe^{iHt}$. The full set of errors that we have to adapt to is huge, including not only the local errors but also highly non-local ones as well. Moreover, without a method for extracting entropy from the system throughout the simulation, these techniques can never be scalable \cite{Pastawski2009}; we simply envisage that they permit larger, more accurate, simulations than in their absence. We report on one special case for which an error corrected simulation is possible: perfect quantum state transfer. This shows conceptually that additional robustness can be imbued upon a quantum simulation, and may provide insight that benefits future studies.

The use of spin chains for transferring a quantum state was first proposed by Bose \cite{Bose2003}, and refined for perfect action in \cite{Christandl2004,kay2010-a}. They are intended to reduce the experimental demands of an essential component of the quantum computer -- the transport of quantum states between distant locations. The concept requires the design, pre-manufacture, and testing of a device that is made from the same technology as the rest of the quantum computer and has a single, fixed function (although `routing' may be possible under certain assumptions \cite{kay2011-a,pemberton-ross2011}). Conceptually, this enables one to expend a lot of effort on making the device as accurately as possible to minimise errors -- since direct control is not required of any of the spins in the chain except for the first and last, all the rest could (in principle) be isolated from the environment. Furthermore, it should be comparatively simple to experimentally realise this protocol \cite{perez-leija2013,weimann2014} (although one has to add remarkably little in order to regain the full power of quantum computation \cite{kay2010}).

Inevitably, errors will arise in both the manufacturing process, and as noise during the transport \cite{perez-leija2013,weimann2014}. How are we to surmount such obstacles? If we allow access to a few sites at the beginning and end of a long chain, then there is an elegant solution to the task of finding the optimal encoding across those spins \cite{haselgrove2005} in the presence of (time independent) identified manufacturing defects. Alternatively, multiple parallel (non-identical) chains may be used, at the cost of a heralded, but non-deterministic arrival \cite{burgarth2005}. Far less is known about dynamically occurring noise, with \cite{marletto2012} imposing that errors only occur at a restricted number of positions and times, and no true error correction protocol (for unidentified errors) is known. 

In this paper, we identify the equivalent of `local errors' for a spin chain and investigate error correcting codes that can correct for the presence of a small number of these errors. In the first result of its kind, we show that by modifying their error correction procedure, the Calderbank-Shor-Steane (CSS) codes \cite{calderbank1996,steane1996} can be used to encode an unknown quantum state into a few sites at the beginning of a long chain, and decoded at the opposite end, thereby enabling high quality transport of a state in the presence of sufficiently low error rates. This study is of significant experimental relevance -- the free-fermion type Hamiltonians which we study have broad experimental feasibility \cite{Blais2004,majer2007,Bialczak2010,Plantenberg2007,Garcia-Ripoll2003}, and this work shows how they can tolerate (i) perturbative errors in the intended coupling strengths (imperfect manufacture), (ii) imperfect timing of the state transfer protocol, and (iii) local noise which is dominated by one particular type of error.

\section{Noisy Transfer Chains}

We consider a chain of $N$ qubits with nearest-neighbour couplings. The Hamiltonian has the XX form
$$
H=\half\sum_{n=1}^NB_nZ_n+\half\sum_{n=1}^{N-1}J_n(X_nX_{n+1}+Y_nY_{n+1}),
$$
where $X_n, Y_n$ and $Z_n$ indicate the standard Pauli $X, Y$ and $Z$ matrices respectively, applied to site $n$ (and identity elsewhere). Such Hamiltonians are particularly appropriate for superconducting qubits \cite{Blais2004,majer2007,Bialczak2010}. The coupling strengths can be selected in many different ways to achieve perfect transfer \cite{karbach2005}, but the details are irrelevant as our constructions are universally applicable -- the same error correcting code and correction procedure can be used on any such chain. Equivalent constructions can be made for other one-dimensional free-fermion models, such as the transverse Ising model \cite{difranco2008,kay2010-a}, which is also experimentally relevant \cite{Plantenberg2007,Garcia-Ripoll2003,Islam2011}, but do not apply to the Heisenberg model. For our purposes, it is sufficient to know that there exists a time $t_0$ such that
\begin{equation}
e^{-iHt_0}\ket{1}\ket{0}^{\otimes (N-1)}=e^{i\phi}\ket{0}^{\otimes (N-1)}\ket{1}	\label{eqn:PST}
\end{equation}
for some known phase $\phi$ \cite{kay2010-a}. Here, the $N$-fold tensor product represents the states of consecutive spins on the chain; the first is the input spin and the last is the output spin of the state transfer process. Eq.\ (\ref{eqn:PST}) imposes that for any arbitrary initial state $\ket{\Psi_I}$ of the $N$ qubits, after evolution for time $t_0$, this state is mirror inverted about the centre of the chain, up to the application of controlled-phase gates between every pair of qubits \cite{kay2010-a,albanese2004}.  However, if a state on a block of spins (such as at either end of the chain) has a fixed parity of excitations (number of $\ket{1}$s), then the controlled-phase gates cannot cause that block to become entangled with the rest of the system. This observation has previously been used to avoid the initialisation of any part of the spin chain except where the state is input \cite{kay2007,difranco2008,kay2010-a}. We seek an encoding for an unknown quantum state on the first $M$ spins of a spin chain such that, after time $t_0$, the input state can be recovered from the $M$ spins at the opposite end of the chain (the decoding region). In the absence of noise, it is sufficient to encode in a state of fixed parity of excitation number and use a perfect transfer chain.

Inspired by the Jordan-Wigner transformation, we introduce the Majorana fermions
$$
c_n=Z_1Z_2\ldots Z_{n-1}X_n\qquad c_{n+N}=Z_1\ldots Z_{n-1}Y_n,
$$
whose time evolution can be written as
\begin{equation}
c_n(t)=e^{-iHt}c_ne^{iHt}=\sum_{m=1}^{2N}\bra{m}e^{-iht}\ket{n}c_m
\label{eqn:evolve}
\end{equation}
where
$$
h=-Y\otimes H_1
$$
is a $2N\times 2N$ matrix describing the coupling of the $2N$ fermion modes (the tensor product between Pauli $Y$ and the matrix $H_1$ is merely a matrix construction and reflects no correspondence with a physical division of subsets of qubits), and
$$
H_1=\sum_{n=1}^NB_n\proj{n}+\sum_{n=1}^{N-1}J_n(\ket{n}\bra{n+1}+\ket{n+1}\bra{n})
$$
is the Hamiltonian $H$, restricted to the first excitation subspace. The key to this description is that the fermions $c_n$ evolve independently of one another (one only has to be careful of the ordering of the operators, which contributes the afore-mentioned controlled-phase gates).

The fermions $c_n$ form a basis that, in principle, any error (such as $X_n$) could be described in terms of. However, certain errors, such as local phase errors are described by only a pair of fermions: $Z_n=-ic_nc_{N+n}$. So, if a phase error were to occur on any spin at any time $t$, then by Eq.\ (\ref{eqn:evolve}), at the state transfer time $t_0$ there would still only be two fermions present in the system, and the errors on the output region would consist of no more than two operators of the form 
\begin{eqnarray*}
&Z_{N+1-M}Z_{N+2-M}\ldots Z_{N+k-1-M}X_{N+k-M}&	\\
&Z_{N+1-M}Z_{N+2-M}\ldots Z_{N+k-1-M}Y_{N+k-M}&
\end{eqnarray*}
where $k\in\{1,\ldots M\}$. These are clearly not the single site operators that standard error correcting codes are designed to combat. However, observe that whatever error occurs: (I) There are no more than 2 bit flips ($X$ or $Y$) and, (II) on a site $p$ where there is no bit-flip, there is a $Z$ error only if there are an odd number of bit-flip errors on the sites $p+1$ to $N$. 

\section{CSS codes} The CSS codes \cite{calderbank1996,steane1996} constitute the first known examples of a quantum error correcting code. An $[[n,k,d]]$ code comprises $n$ physical qubits, encoding $k$ logical qubits and is capable of correcting for any $\lfloor (d-1)/2 \rfloor$ single-qubit errors ($X$, $Y$ or $Z$). They work by combining two different classical codes: one for $X$ errors and the other for $Z$ errors. After an encoded state has been exposed to noise, the original state can be reconstructed by first performing a syndrome extraction, in which the location of errors is written onto some ancilla qubits, and then error correction, in which the ancillas are measured and the detected errors are inverted, for each type of error ($X$ or $Z$) in turn.

Assume that we have encoded into a $[[M,1,d]]$ CSS code of distance $d\geq 5$, with the additional constraint that all the stabilizers and logical operators of the code commute with $Z^{\otimes M}$. The perfect mirroring property of the perfect transfer system ensures that the code arrives perfectly on the decoding region at time $t_0$, up to the possible controlled-phases that are globally applied. The commutativity of the code with $Z^{\otimes M}$ ensures that the code space has a fixed parity of excitation number, and hence the rest of the chain can be initialised in an arbitrary state and not get entangled with the decoding region due to these gates. The internal controlled-phase gates are removed by applying controlled-phase gates between all pairs of qubits $N+1-M$ to $N$ \footnote{Alternatively, one could simply update the stabilizers of the code.}, returning the original code, but updating the errors to
\begin{eqnarray*}
&X_{N+k-M}Z_{N+k+1-M}Z_{N+k+2-M}\ldots Z_N&	\\
&Y_{N+k-M}Z_{N+k+1-M}Z_{N+k+2-M}\ldots Z_N.&
\end{eqnarray*}
By performing standard syndrome extraction for the $X$-type errors, we can detect any pair of bit-flip errors. Hence, by observation (I) we can detect the location of any fermionic operators that lie in the decoding region (with the exception of $c_nc_{N+n}$). This is followed up by an error correction step. Obviously, we should apply $X$ rotations to the detected error locations as normal. However, by observation (II) we also apply a $Z$ rotation to any spin $n$ that has an odd number of detected bit flips on sites $N+1-M$ to $n-1$. This change in the procedure negates some of the fault-tolerant properties of CSS codes \cite{aliferis2006}, but these are irrelevant here.

The only errors that we haven't corrected for are either the $c_nc_{N+n}$ type (a $Z$ on a single site), or the distinction between $X$ and $Y$ operators, which were corrected as if they were $X$. Thus, the remaining errors consist of up to two $Z$ errors, which can be detected and corrected using the $Z$ part of the CSS code in the standard manner. The net result is the perfect correction of any single 2-fermion error (e.g. $Z$) that occurs at any time during the evolution of the system, a far stronger result than the identified noise of \cite{marletto2012}, while by no means contradicting studies of information propagation in more noisy scenarios \cite{burrell2009}.

More generally, by selecting an $[[M,1,2k+1]]$ CSS code to encode into (whose stabilizers commute with $Z^{\otimes M}$), we can correct for up to $k$ Majorana fermions arriving in the output region. This requires input and output regions of size $O(k^2)$. While such codes are not so well studied,  $[[d^2,1,d]]$ Shor codes can easily be constructed: divide the $d^2$ qubits into blocks of $d$ and define logical qubits $\ket{0}_{L1}=\ket{+}^{\otimes d}$ and $\ket{1}_{L1}=\ket{-}^{\otimes d}$ (Hadamard-rotated majority vote for $Z$ errors, distance $d$, where $\ket{\pm}=(\ket{0}\pm\ket{1})/\sqrt{2}$), then take those and form them into a standard majority vote (i.e.\ tuned for $X$ errors) $\ket{0}_{L2}=\ket{0}_{L1}^{\otimes d}$ and $\ket{1}_{L2}=\ket{1}_{L1}^{\otimes d}$. This has a fixed parity of excitation number provided $d$ is even. However, better codes probably exist -- in the case of $d=5$, a $[[19,1,5]]$ code has been found \footnote{P.\ M\'ajek, Diploma Thesis, Comenius University (2005)}, although this does not have the necessary excitation parity condition. Indeed, the explicit example that we have simulated in the next section suggests that asymmetric quantum error correcting codes could become objects of particular interest \cite{ioffe2007}.

\subsection{Minimal Working Example}

\begin{figure}[!tbp]
\begin{center}
\includegraphics[width=0.42\textwidth]{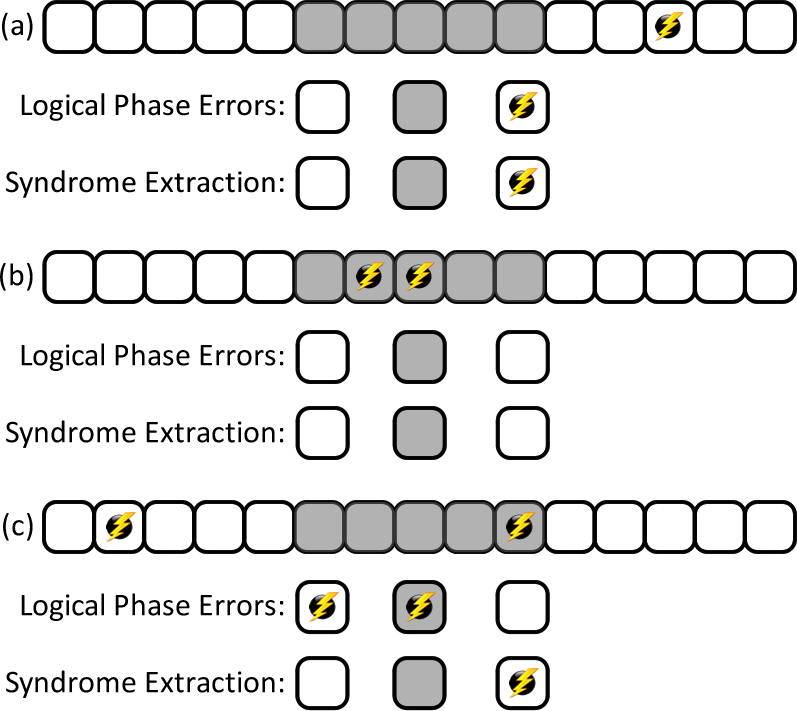}
\vspace{-0.5cm}
\end{center}
\caption{(Color online) An error correcting code of 15 qubits made up of 3 logical qubits (grouped by shading) in a $Z$-error correcting code, each composed of a 5-qubit $X$-error correcting code. Two bit-flips, and the trailing $Z$ errors, have been detected and corrected. Up to two $Z$ errors remain, located on the sites of the detected $X$ errors. A single error is easily detected and corrected (a). A pair of errors within the same logical qubit cancel each other (b). Two errors on different logical qubits yield a syndrome that does not match the locations of the $X$ errors, flagging the need to correct for two $Z$ errors.} \label{fig:min}
\vspace{-0.5cm}
\end{figure}

As presented, the minimal size of the encoding and decoding regions is 36 qubits, rendering simulation of the full Hilbert space a serious challenge. In order to numerically verify the presented results, we made a number of simplifications. Firstly, we limited the length of the chain to the size of the encoding region ($M=N$), and evolved for twice the state transfer time, thereby creating a perfect revival of the original state, which also removes the requirement for the code to commute with $Z^{\otimes M}$. Secondly, we made the observation that although up to two $Z$ errors could occur, it is only necessary to use an error correcting code that corrects for a single $Z$ error. This is because there are two possible cases for what happens: either no bit flip error is detected (which means the $Z$ error has propagated to a single $Z$ error on some site), or two bit flip errors are detected (which means that there will be up to two $Z$ errors on those two sites). Obviously, the only case that a one-$Z$-error correcting code could not implicitly deal with is the instance in which there are exactly two $Z$ errors, Fig.\ \ref{fig:min}. However, we know that if a one-$Z$-error correcting code is comprised of 3 qubits, and there are errors on two of the qubits, the syndrome measurement detects an error on the \emph{other} bit, Fig.\ \ref{fig:min}(c). Thus, if we compare this with the syndrome information for the $X$ error, we can implement the additional rule ``if a $Z$ error is detected on a logical qubit on which no $X$ error was detected, then there were actually $Z$ errors on the other two logical bits instead''. As such, the Shor-like construction reduces to 15 qubits, rendering simulation more feasible. We have implemented this \footnote{A.\ Kay (2016): full\_test.nb. figshare. doi:\detokenize{10.6084/m9.figshare.2070181}} using the standard coupling configuration for perfect transfer, $J_n=\sqrt{n(N-n)}$ and $B_n=0$ \cite{Christandl2004,kay2010-a}. For 1024 random samples of a single $Z$ error (chosen randomly to occur at any time during the transfer time, and to be applied at any random site on the chain), every instance was corrected perfectly.

\section{More Realistic Noise}

While a fixed number of errors simplifies the pedagogy, of more practical interest is the case of a fixed per-qubit error rate. Assume that there is a probability per qubit and per unit time of $\gamma$ that a two-fermion error occurs on the spin chain. For instance, phase errors can be described using the Master Equation
$$
\frac{d\rho}{dt}=-i[H,\rho]-N\gamma\rho+\gamma\sum_{n=1}^NZ_n\rho Z_n.
$$
During the time $t_0$, one would therefore expect an average of $2\gamma Nt_0$ fermionic errors to afflict the chain. However, we anticipate that only $O(\gamma Mt_0)$ of these are located on the decoding region at the moment of error correction. To formalise this, consider the intended transformation under the perfect state transfer Hamiltonian of $c_n\mapsto c_{N+1-n}$ and $c_{N+n}\mapsto c_{2N+1-n}$, i.e.\ we want to keep track of errors in the mode $e^{-iHt}c_ne^{iHt}=e^{iHt}c_{N+1-n}e^{-iHt}$ for $n\leq N$. Thus, if we evaluate
$$
\chi_n=\Tr(\rho e^{-iHt}c_ne^{iHt})=\Tr(\tilde\rho c_{N+1-n}),
$$
where $\tilde\rho=e^{-iHt}\rho e^{iHt}$ is the density matrix in the interaction picture, we can evaluate the derivative as
\begin{multline*}
\frac{d\chi_n}{dt}=-N\gamma\chi_n+\\
\gamma\sum_{m=1}^N\Tr\left(\tilde\rho e^{-iHt}Z_me^{iHt}c_{N+1-n}e^{-iHt}Z_me^{iHt}\right).
\end{multline*}
Using Eq.\ (\ref{eqn:evolve}), we rewrite
\begin{eqnarray*}
Z_me^{iHt}c_ne^{-iHt}Z_m&=&e^{iHt}c_ne^{-iHt}-2c_m\bra{m}e^{iht}\ket{n}	\\
&&-2c_{N+m}\bra{N+m}e^{iht}\ket{n}
\end{eqnarray*}
such that
\begin{eqnarray*}
\frac{d\chi_n}{dt}&=&-2\gamma\Tr\left(\tilde\rho e^{-iHt}\sum_{m=1}^{2N}\bra{m}e^{iht}\ket{N+1-n}c_me^{iHt}\right)	\\
&=&-2\gamma\chi_n,
\end{eqnarray*}
using Eq.\ (\ref{eqn:evolve}) again. This leaves a final solution of $\chi_n(t)=e^{-2\gamma t}\chi_n(0)$. We interpret this as a probability of $p=\half(1-e^{-2\gamma t})$ of each fermionic mode having an error. If $\gamma t_0\ll 1$, then the error probability is approximately $p=\gamma t_0$ per fermionic error. However, these error modes do not occur independently of one another. Provided the expected number of errors $\sim2\gamma Mt_0$ is smaller than the number of errors that the code can correct for ($\sim\sqrt{M}/2$), the code can be useful. Thus, we require $\gamma t_0\sim1/\sqrt{M}$ and, furthermore, $t_0$ scales with at least $N$ if the maximum coupling strength of a chain is bounded \cite{yung2006}. We therefore envisage this being applied in finite length chains where the parameters can be judiciously chosen to be effective, thereby providing regular repetitions of error correction if transfer is required over greater distances.

\section{Timing and Manufacturing Errors}

\begin{figure}[!tbp]
\begin{center}
\includegraphics[width=0.42\textwidth]{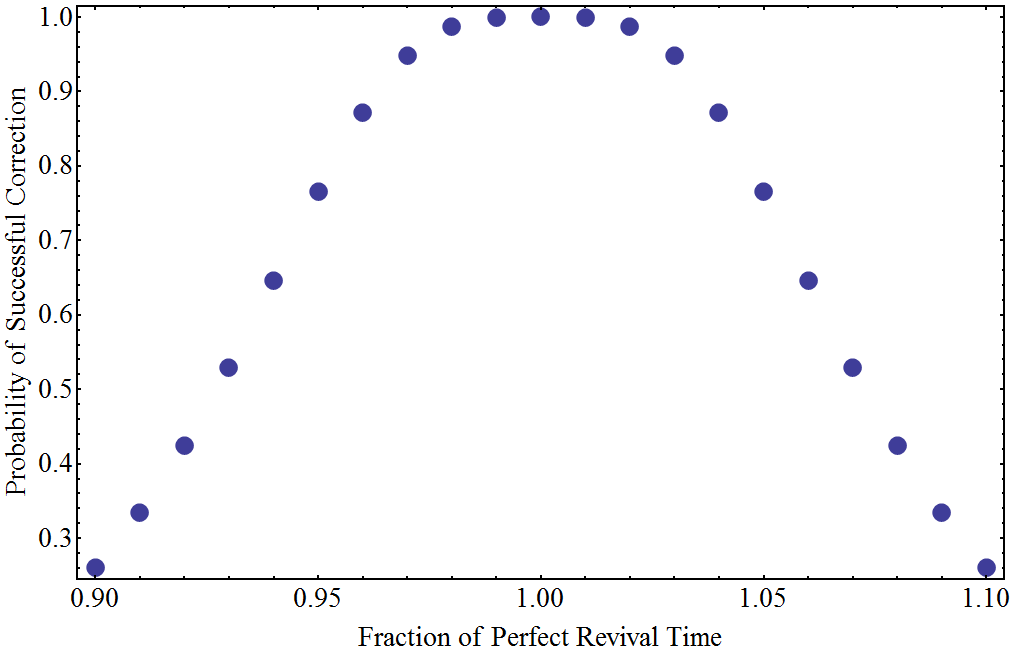}
\vspace{-0.5cm}
\end{center}
\caption{In the minimal working example, we test faults due to timing errors, assessing the probability of successful corrections (by tracking all possible syndrome measurements).} \label{fig:timing}
\vspace{-0.5cm}
\end{figure}

So far, we have shown that a small number of fermionic errors can be corrected for, and we have ascribed their appearance to noise in a particular basis. However, there are two other important mechanisms that can be described in this way. The first is a timing error -- rather than removing the arriving state from the spin chain at time $t_0$, we accidentally do so at the time $t_0+\delta t$ \cite{kay2006}. Instead of the initial state $\ket{\Psi_I}$ evolving to the target state $\ket{\Psi_T}=e^{-iHt_0}\ket{\Psi_I}$, it acquires an error $e^{-iH\delta t}$. Expanding this for small $\delta t$ (requiring $\delta t\lambda_{\max}\ll 1$, where $\lambda_{\max}$ is the largest singular value of $H_1$) yields an expansion in even powers of the fermionic operators, i.e.\ larger numbers of errors are strongly suppressed such that error correction succeeds with high probability. In Fig.\ \ref{fig:timing}, we show how the minimal working example successfully corrects the majority of cases for a small timing error. Note that, unlike previous treatments such as in \cite{kay2006}, here we evaluate the probability that the state arrives perfectly, not the overlap between the input and output states.

\begin{figure}[!tbp]
\begin{center}
\includegraphics[width=0.42\textwidth]{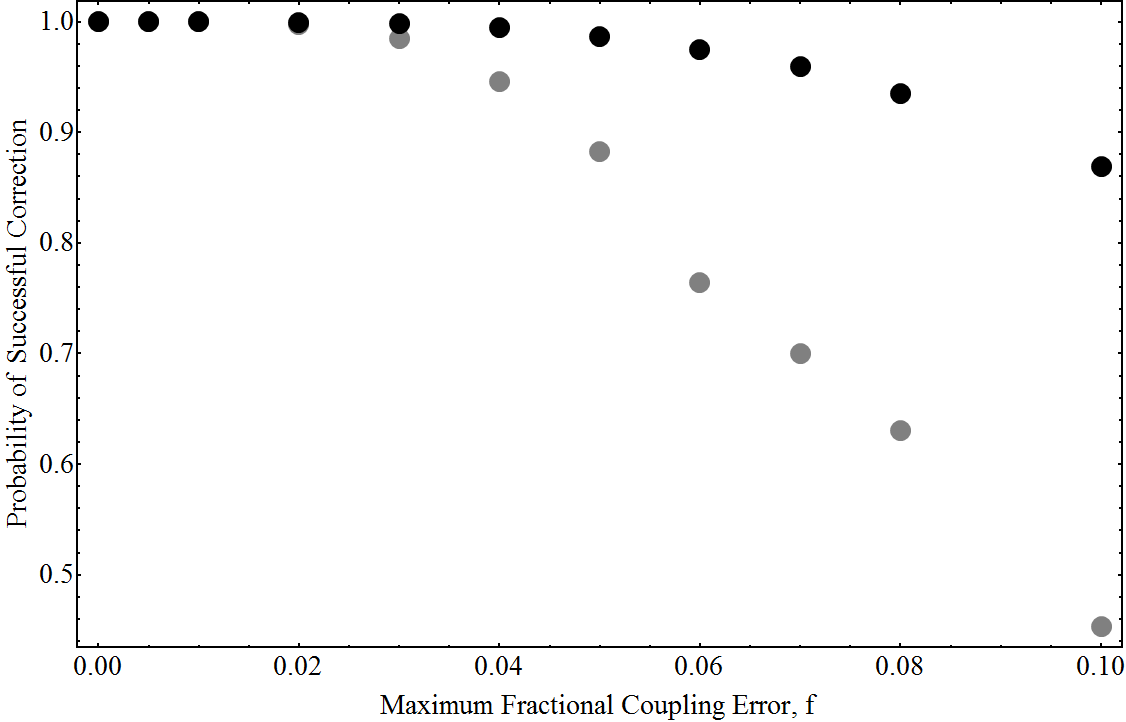}
\vspace{-0.5cm}
\end{center}
\caption{In the minimal working example, we test faults due to coupling errors, altering each uniformly at random in the range $J_n(1-f)$ to $J_n(1+f)$, assessing the probability of successful corrections and averaging over 1000 different instances (black) or the minimum success probability (grey).} \label{fig:coupling}
\vspace{-0.5cm}
\end{figure}

Similarly, were we to imperfectly manufacture the target coupling strengths and magnetic fields in $H$, then the perturbation $V$ (which could be time dependent) is quadratic in fermions. Now the error in evolution can be described by $e^{-i(H+V)t_0}e^{iHt_0}\ket{\Psi_T}$, and expanded in powers of $V$ as
$$
e^{-i(H+V)t_0}e^{iHt_0}=\identity+it_0\int_0^{t_0}e^{-i Ht}Ve^{i Ht}dt+O(t_0^2V^2).
$$
Since the action of $e^{-iHt}$ preserves the number of fermions, this represents an expansion in even powers of fermionic operators. 
Thus, provided $\zeta_{\max} t_0\ll 1$ where $\zeta_{\max}$ is the largest singular value of the first excitation subspace of $V$, this describes that with high probability the number of fermionic errors is small, and these can be corrected for via our error correction procedure. The efficacy of this procedure is tested for the minimal working example and depicted in Fig.\ \ref{fig:coupling}. Of course, a better test would be to utilise state transfer over a greater chain length to avoid possible confusion with localisation due to the errors in the system.

\section{Conclusions} In spite of the fact that CSS codes were designed for application in the scenario of local, independent, errors, we have shown that they can be retasked to correct for the massively correlated errors that typically arise due to the intrinsic Hamiltonian dynamics of a spin chain, such as those intended to perform the perfect quantum state transfer of spin chains. This in turn shows that additional robustness against imperfections can be embedded into a fixed-function Hamiltonian evolution such as an analog quantum simulation, even if it was dependent on some very specific features of the Hamiltonian (the free-fermion structure).

The major drawback of encoding in this way is that a block encoding of size $M$ can tolerate at most $O(\sqrt{M})$ errors, but a constant per-qubit error rate would require $O(M)$ to be tolerated, imposing the requirement for rather small error rates. A future direction would be to try and improve this through better choice of error correcting code. We are given heart by the surface code \cite{kitaev2003} -- this is a CSS code and, although its distance is short, it is the case that in the presence of local noise, it is highly unlikely that those short error strings that cause failure of the code arise. Indeed, the surface code has an error correcting threshold consisting of a finite per-qubit error rate, exactly as we desire \cite{dennis2002,kay2014}. However, more work would be required -- the errors in the present model are not independent. Provided these correlations are local, it is likely that they could be tolerated, with a mapping from the spin chain onto the surface code that preserves the locality of the errors.

The error correction succeeds provided the error operators are well described in terms of a small number of fermionic operators. This includes terms such as $Z$, $X\otimes X$ and $Y\otimes Y$, as well as Hamiltonian perturbations and timing errors. However, it does not include bit-flip errors -- a bit-flip on spin $n$ requires $2n-1$ Majorana fermions to describe it. It remains an interesting question for the future whether either a different error correction strategy, or a different class of Hamiltonians, permits error correction of all local errors. 

\bibliography{../../../References}

\begin{thebibliography}{41}%
\makeatletter
\providecommand \@ifxundefined [1]{%
 \@ifx{#1\undefined}
}%
\providecommand \@ifnum [1]{%
 \ifnum #1\expandafter \@firstoftwo
 \else \expandafter \@secondoftwo
 \fi
}%
\providecommand \@ifx [1]{%
 \ifx #1\expandafter \@firstoftwo
 \else \expandafter \@secondoftwo
 \fi
}%
\providecommand \natexlab [1]{#1}%
\providecommand \enquote  [1]{``#1''}%
\providecommand \bibnamefont  [1]{#1}%
\providecommand \bibfnamefont [1]{#1}%
\providecommand \citenamefont [1]{#1}%
\providecommand \href@noop [0]{\@secondoftwo}%
\providecommand \href [0]{\begingroup \@sanitize@url \@href}%
\providecommand \@href[1]{\@@startlink{#1}\@@href}%
\providecommand \@@href[1]{\endgroup#1\@@endlink}%
\providecommand \@sanitize@url [0]{\catcode `\\12\catcode `\$12\catcode
  `\&12\catcode `\#12\catcode `\^12\catcode `\_12\catcode `\%12\relax}%
\providecommand \@@startlink[1]{}%
\providecommand \@@endlink[0]{}%
\providecommand \url  [0]{\begingroup\@sanitize@url \@url }%
\providecommand \@url [1]{\endgroup\@href {#1}{\urlprefix }}%
\providecommand \urlprefix  [0]{URL }%
\providecommand \Eprint [0]{\href }%
\providecommand \doibase [0]{http://dx.doi.org/}%
\providecommand \selectlanguage [0]{\@gobble}%
\providecommand \bibinfo  [0]{\@secondoftwo}%
\providecommand \bibfield  [0]{\@secondoftwo}%
\providecommand \translation [1]{[#1]}%
\providecommand \BibitemOpen [0]{}%
\providecommand \bibitemStop [0]{}%
\providecommand \bibitemNoStop [0]{.\EOS\space}%
\providecommand \EOS [0]{\spacefactor3000\relax}%
\providecommand \BibitemShut  [1]{\csname bibitem#1\endcsname}%
\let\auto@bib@innerbib\@empty
\bibitem [{\citenamefont {Somaroo}\ \emph {et~al.}(1999)\citenamefont
  {Somaroo}, \citenamefont {Tseng}, \citenamefont {Havel}, \citenamefont
  {Laflamme},\ and\ \citenamefont {Cory}}]{Somaroo1999}%
  \BibitemOpen
  \bibfield  {author} {\bibinfo {author} {\bibfnamefont {S.}~\bibnamefont
  {Somaroo}}, \bibinfo {author} {\bibfnamefont {C.~H.}\ \bibnamefont {Tseng}},
  \bibinfo {author} {\bibfnamefont {T.~F.}\ \bibnamefont {Havel}}, \bibinfo
  {author} {\bibfnamefont {R.}~\bibnamefont {Laflamme}}, \ and\ \bibinfo
  {author} {\bibfnamefont {D.~G.}\ \bibnamefont {Cory}},\ }\href {\doibase
  10.1103/PhysRevLett.82.5381} {\bibfield  {journal} {\bibinfo  {journal}
  {Physical Review Letters}\ }\textbf {\bibinfo {volume} {82}},\ \bibinfo
  {pages} {5381} (\bibinfo {year} {1999})}\BibitemShut {NoStop}%
\bibitem [{\citenamefont {Friedenauer}\ \emph {et~al.}(2008)\citenamefont
  {Friedenauer}, \citenamefont {Schmitz}, \citenamefont {Glueckert},
  \citenamefont {Porras},\ and\ \citenamefont {Schaetz}}]{Friedenauer2008}%
  \BibitemOpen
  \bibfield  {author} {\bibinfo {author} {\bibfnamefont {A.}~\bibnamefont
  {Friedenauer}}, \bibinfo {author} {\bibfnamefont {H.}~\bibnamefont
  {Schmitz}}, \bibinfo {author} {\bibfnamefont {J.~T.}\ \bibnamefont
  {Glueckert}}, \bibinfo {author} {\bibfnamefont {D.}~\bibnamefont {Porras}}, \
  and\ \bibinfo {author} {\bibfnamefont {T.}~\bibnamefont {Schaetz}},\ }\href
  {\doibase 10.1038/nphys1032} {\bibfield  {journal} {\bibinfo  {journal}
  {Nature Physics}\ }\textbf {\bibinfo {volume} {4}},\ \bibinfo {pages} {757}
  (\bibinfo {year} {2008})}\BibitemShut {NoStop}%
\bibitem [{\citenamefont {Gerritsma}\ \emph {et~al.}(2010)\citenamefont
  {Gerritsma}, \citenamefont {Kirchmair}, \citenamefont {Z\"{a}hringer},
  \citenamefont {Solano}, \citenamefont {Blatt},\ and\ \citenamefont
  {Roos}}]{Gerritsma2010}%
  \BibitemOpen
  \bibfield  {author} {\bibinfo {author} {\bibfnamefont {R.}~\bibnamefont
  {Gerritsma}}, \bibinfo {author} {\bibfnamefont {G.}~\bibnamefont
  {Kirchmair}}, \bibinfo {author} {\bibfnamefont {F.}~\bibnamefont
  {Z\"{a}hringer}}, \bibinfo {author} {\bibfnamefont {E.}~\bibnamefont
  {Solano}}, \bibinfo {author} {\bibfnamefont {R.}~\bibnamefont {Blatt}}, \
  and\ \bibinfo {author} {\bibfnamefont {C.~F.}\ \bibnamefont {Roos}},\ }\href
  {\doibase 10.1038/nature08688} {\bibfield  {journal} {\bibinfo  {journal}
  {Nature}\ }\textbf {\bibinfo {volume} {463}},\ \bibinfo {pages} {68}
  (\bibinfo {year} {2010})}\BibitemShut {NoStop}%
\bibitem [{\citenamefont {Kim}\ \emph {et~al.}(2010)\citenamefont {Kim},
  \citenamefont {Chang}, \citenamefont {Korenblit}, \citenamefont {Islam},
  \citenamefont {Edwards}, \citenamefont {Freericks}, \citenamefont {Lin},
  \citenamefont {Duan},\ and\ \citenamefont {Monroe}}]{kim2010}%
  \BibitemOpen
  \bibfield  {author} {\bibinfo {author} {\bibfnamefont {K.}~\bibnamefont
  {Kim}}, \bibinfo {author} {\bibfnamefont {M.}~\bibnamefont {Chang}}, \bibinfo
  {author} {\bibfnamefont {S.}~\bibnamefont {Korenblit}}, \bibinfo {author}
  {\bibfnamefont {R.}~\bibnamefont {Islam}}, \bibinfo {author} {\bibfnamefont
  {E.~E.}\ \bibnamefont {Edwards}}, \bibinfo {author} {\bibfnamefont {J.~K.}\
  \bibnamefont {Freericks}}, \bibinfo {author} {\bibfnamefont {G.}~\bibnamefont
  {Lin}}, \bibinfo {author} {\bibfnamefont {L.}~\bibnamefont {Duan}}, \ and\
  \bibinfo {author} {\bibfnamefont {C.}~\bibnamefont {Monroe}},\ }\href
  {\doibase 10.1038/nature09071} {\bibfield  {journal} {\bibinfo  {journal}
  {Nature}\ }\textbf {\bibinfo {volume} {465}},\ \bibinfo {pages} {590}
  (\bibinfo {year} {2010})}\BibitemShut {NoStop}%
\bibitem [{\citenamefont {Islam}\ \emph {et~al.}(2011)\citenamefont {Islam},
  \citenamefont {Edwards}, \citenamefont {Kim}, \citenamefont {Korenblit},
  \citenamefont {Noh}, \citenamefont {Carmichael}, \citenamefont {Lin},
  \citenamefont {Duan}, \citenamefont {Joseph~Wang}, \citenamefont
  {Freericks},\ and\ \citenamefont {Monroe}}]{Islam2011}%
  \BibitemOpen
  \bibfield  {author} {\bibinfo {author} {\bibfnamefont {R.}~\bibnamefont
  {Islam}}, \bibinfo {author} {\bibfnamefont {E.~E.}\ \bibnamefont {Edwards}},
  \bibinfo {author} {\bibfnamefont {K.}~\bibnamefont {Kim}}, \bibinfo {author}
  {\bibfnamefont {S.}~\bibnamefont {Korenblit}}, \bibinfo {author}
  {\bibfnamefont {C.}~\bibnamefont {Noh}}, \bibinfo {author} {\bibfnamefont
  {H.}~\bibnamefont {Carmichael}}, \bibinfo {author} {\bibfnamefont
  {G.}~\bibnamefont {Lin}}, \bibinfo {author} {\bibfnamefont {L.}~\bibnamefont
  {Duan}}, \bibinfo {author} {\bibfnamefont {C.}~\bibnamefont {Joseph~Wang}},
  \bibinfo {author} {\bibfnamefont {J.~K.}\ \bibnamefont {Freericks}}, \ and\
  \bibinfo {author} {\bibfnamefont {C.}~\bibnamefont {Monroe}},\ }\href
  {\doibase 10.1038/ncomms1374} {\bibfield  {journal} {\bibinfo  {journal}
  {Nature Communications}\ }\textbf {\bibinfo {volume} {2}},\ \bibinfo {pages}
  {377} (\bibinfo {year} {2011})}\BibitemShut {NoStop}%
\bibitem [{\citenamefont {Lanyon}\ \emph {et~al.}(2011)\citenamefont {Lanyon},
  \citenamefont {Hempel}, \citenamefont {Nigg}, \citenamefont {M\"{u}ller},
  \citenamefont {Gerritsma}, \citenamefont {Z\"{a}hringer}, \citenamefont
  {Schindler}, \citenamefont {Barreiro}, \citenamefont {Rambach}, \citenamefont
  {Kirchmair}, \citenamefont {Hennrich}, \citenamefont {Zoller}, \citenamefont
  {Blatt},\ and\ \citenamefont {Roos}}]{lanyon2011}%
  \BibitemOpen
  \bibfield  {author} {\bibinfo {author} {\bibfnamefont {B.~P.}\ \bibnamefont
  {Lanyon}}, \bibinfo {author} {\bibfnamefont {C.}~\bibnamefont {Hempel}},
  \bibinfo {author} {\bibfnamefont {D.}~\bibnamefont {Nigg}}, \bibinfo {author}
  {\bibfnamefont {M.}~\bibnamefont {M\"{u}ller}}, \bibinfo {author}
  {\bibfnamefont {R.}~\bibnamefont {Gerritsma}}, \bibinfo {author}
  {\bibfnamefont {F.}~\bibnamefont {Z\"{a}hringer}}, \bibinfo {author}
  {\bibfnamefont {P.}~\bibnamefont {Schindler}}, \bibinfo {author}
  {\bibfnamefont {J.~T.}\ \bibnamefont {Barreiro}}, \bibinfo {author}
  {\bibfnamefont {M.}~\bibnamefont {Rambach}}, \bibinfo {author} {\bibfnamefont
  {G.}~\bibnamefont {Kirchmair}}, \bibinfo {author} {\bibfnamefont
  {M.}~\bibnamefont {Hennrich}}, \bibinfo {author} {\bibfnamefont
  {P.}~\bibnamefont {Zoller}}, \bibinfo {author} {\bibfnamefont
  {R.}~\bibnamefont {Blatt}}, \ and\ \bibinfo {author} {\bibfnamefont {C.~F.}\
  \bibnamefont {Roos}},\ }\href {\doibase 10.1126/science.1208001} {\bibfield
  {journal} {\bibinfo  {journal} {Science}\ }\textbf {\bibinfo {volume}
  {334}},\ \bibinfo {pages} {57} (\bibinfo {year} {2011})},\ \bibinfo {note}
  {{PMID:} 21885735}\BibitemShut {NoStop}%
\bibitem [{\citenamefont {Jones}\ \emph {et~al.}(2012)\citenamefont {Jones},
  \citenamefont {Whitfield}, \citenamefont {{McMahon}}, \citenamefont {Yung},
  \citenamefont {Meter}, \citenamefont {{Aspuru-Guzik}},\ and\ \citenamefont
  {Yamamoto}}]{jones2012}%
  \BibitemOpen
  \bibfield  {author} {\bibinfo {author} {\bibfnamefont {N.~C.}\ \bibnamefont
  {Jones}}, \bibinfo {author} {\bibfnamefont {J.~D.}\ \bibnamefont
  {Whitfield}}, \bibinfo {author} {\bibfnamefont {P.~L.}\ \bibnamefont
  {{McMahon}}}, \bibinfo {author} {\bibfnamefont {M.}~\bibnamefont {Yung}},
  \bibinfo {author} {\bibfnamefont {R.~V.}\ \bibnamefont {Meter}}, \bibinfo
  {author} {\bibfnamefont {A.}~\bibnamefont {{Aspuru-Guzik}}}, \ and\ \bibinfo
  {author} {\bibfnamefont {Y.}~\bibnamefont {Yamamoto}},\ }\href {\doibase
  10.1088/1367-2630/14/11/115023} {\bibfield  {journal} {\bibinfo  {journal}
  {New Journal of Physics}\ }\textbf {\bibinfo {volume} {14}},\ \bibinfo
  {pages} {115023} (\bibinfo {year} {2012})}\BibitemShut {NoStop}%
\bibitem [{\citenamefont {Pastawski}\ \emph {et~al.}(2009)\citenamefont
  {Pastawski}, \citenamefont {Kay}, \citenamefont {Schuch},\ and\ \citenamefont
  {Cirac}}]{Pastawski2009}%
  \BibitemOpen
  \bibfield  {author} {\bibinfo {author} {\bibfnamefont {F.}~\bibnamefont
  {Pastawski}}, \bibinfo {author} {\bibfnamefont {A.}~\bibnamefont {Kay}},
  \bibinfo {author} {\bibfnamefont {N.}~\bibnamefont {Schuch}}, \ and\ \bibinfo
  {author} {\bibfnamefont {I.}~\bibnamefont {Cirac}},\ }\href {\doibase
  10.1103/PhysRevLett.103.080501} {\bibfield  {journal} {\bibinfo  {journal}
  {Physical Review Letters}\ }\textbf {\bibinfo {volume} {103}},\ \bibinfo
  {pages} {080501} (\bibinfo {year} {2009})}\BibitemShut {NoStop}%
\bibitem [{\citenamefont {Bose}(2003)}]{Bose2003}%
  \BibitemOpen
  \bibfield  {author} {\bibinfo {author} {\bibfnamefont {S.}~\bibnamefont
  {Bose}},\ }\href {\doibase 10.1103/PhysRevLett.91.207901} {\bibfield
  {journal} {\bibinfo  {journal} {Physical Review Letters}\ }\textbf {\bibinfo
  {volume} {91}},\ \bibinfo {pages} {207901} (\bibinfo {year}
  {2003})}\BibitemShut {NoStop}%
\bibitem [{\citenamefont {Christandl}\ \emph {et~al.}(2004)\citenamefont
  {Christandl}, \citenamefont {Datta}, \citenamefont {Ekert},\ and\
  \citenamefont {Landahl}}]{Christandl2004}%
  \BibitemOpen
  \bibfield  {author} {\bibinfo {author} {\bibfnamefont {M.}~\bibnamefont
  {Christandl}}, \bibinfo {author} {\bibfnamefont {N.}~\bibnamefont {Datta}},
  \bibinfo {author} {\bibfnamefont {A.}~\bibnamefont {Ekert}}, \ and\ \bibinfo
  {author} {\bibfnamefont {A.~J.}\ \bibnamefont {Landahl}},\ }\href {\doibase
  10.1103/PhysRevLett.92.187902} {\bibfield  {journal} {\bibinfo  {journal}
  {Physical Review Letters}\ }\textbf {\bibinfo {volume} {92}},\ \bibinfo
  {pages} {187902} (\bibinfo {year} {2004})}\BibitemShut {NoStop}%
\bibitem [{\citenamefont {Kay}(2010)}]{kay2010-a}%
  \BibitemOpen
  \bibfield  {author} {\bibinfo {author} {\bibfnamefont {A.}~\bibnamefont
  {Kay}},\ }\href {http://arxiv.org/abs/0903.4274} {\bibfield  {journal}
  {\bibinfo  {journal} {Int. J. Quantum Inform.}\ }\textbf {\bibinfo {volume}
  {8}},\ \bibinfo {pages} {641} (\bibinfo {year} {2010})}\BibitemShut {NoStop}%
\bibitem [{\citenamefont {Kay}(2011)}]{kay2011-a}%
  \BibitemOpen
  \bibfield  {author} {\bibinfo {author} {\bibfnamefont {A.}~\bibnamefont
  {Kay}},\ }\href {\doibase 10.1103/PhysRevA.84.022337} {\bibfield  {journal}
  {\bibinfo  {journal} {Physical Review A}\ }\textbf {\bibinfo {volume} {84}},\
  \bibinfo {pages} {022337} (\bibinfo {year} {2011})}\BibitemShut {NoStop}%
\bibitem [{\citenamefont {{Pemberton-Ross}}\ and\ \citenamefont
  {Kay}(2011)}]{pemberton-ross2011}%
  \BibitemOpen
  \bibfield  {author} {\bibinfo {author} {\bibfnamefont {P.~J.}\ \bibnamefont
  {{Pemberton-Ross}}}\ and\ \bibinfo {author} {\bibfnamefont {A.}~\bibnamefont
  {Kay}},\ }\href {\doibase 10.1103/PhysRevLett.106.020503} {\bibfield
  {journal} {\bibinfo  {journal} {Physical Review Letters}\ }\textbf {\bibinfo
  {volume} {106}},\ \bibinfo {pages} {020503} (\bibinfo {year}
  {2011})}\BibitemShut {NoStop}%
\bibitem [{\citenamefont {{Perez-Leija}}\ \emph {et~al.}(2013)\citenamefont
  {{Perez-Leija}}, \citenamefont {Keil}, \citenamefont {Kay}, \citenamefont
  {{Moya-Cessa}}, \citenamefont {Nolte}, \citenamefont {Kwek}, \citenamefont
  {{Rodr\'{i}guez-Lara}}, \citenamefont {Szameit},\ and\ \citenamefont
  {Christodoulides}}]{perez-leija2013}%
  \BibitemOpen
  \bibfield  {author} {\bibinfo {author} {\bibfnamefont {A.}~\bibnamefont
  {{Perez-Leija}}}, \bibinfo {author} {\bibfnamefont {R.}~\bibnamefont {Keil}},
  \bibinfo {author} {\bibfnamefont {A.}~\bibnamefont {Kay}}, \bibinfo {author}
  {\bibfnamefont {H.}~\bibnamefont {{Moya-Cessa}}}, \bibinfo {author}
  {\bibfnamefont {S.}~\bibnamefont {Nolte}}, \bibinfo {author} {\bibfnamefont
  {L.}~\bibnamefont {Kwek}}, \bibinfo {author} {\bibfnamefont {B.~M.}\
  \bibnamefont {{Rodr\'{i}guez-Lara}}}, \bibinfo {author} {\bibfnamefont
  {A.}~\bibnamefont {Szameit}}, \ and\ \bibinfo {author} {\bibfnamefont
  {D.~N.}\ \bibnamefont {Christodoulides}},\ }\href {\doibase
  10.1103/PhysRevA.87.012309} {\bibfield  {journal} {\bibinfo  {journal}
  {Physical Review A}\ }\textbf {\bibinfo {volume} {87}},\ \bibinfo {pages}
  {012309} (\bibinfo {year} {2013})}\BibitemShut {NoStop}%
\bibitem [{\citenamefont {Weimann}\ \emph {et~al.}(2014)\citenamefont
  {Weimann}, \citenamefont {Kay}, \citenamefont {Keil}, \citenamefont {Nolte},\
  and\ \citenamefont {Szameit}}]{weimann2014}%
  \BibitemOpen
  \bibfield  {author} {\bibinfo {author} {\bibfnamefont {S.}~\bibnamefont
  {Weimann}}, \bibinfo {author} {\bibfnamefont {A.}~\bibnamefont {Kay}},
  \bibinfo {author} {\bibfnamefont {R.}~\bibnamefont {Keil}}, \bibinfo {author}
  {\bibfnamefont {S.}~\bibnamefont {Nolte}}, \ and\ \bibinfo {author}
  {\bibfnamefont {A.}~\bibnamefont {Szameit}},\ }\href {\doibase
  10.1364/OL.39.000123} {\bibfield  {journal} {\bibinfo  {journal} {Optics
  Letters}\ }\textbf {\bibinfo {volume} {39}},\ \bibinfo {pages} {123}
  (\bibinfo {year} {2014})}\BibitemShut {NoStop}%
\bibitem [{\citenamefont {Kay}\ and\ \citenamefont
  {{Pemberton-Ross}}(2010)}]{kay2010}%
  \BibitemOpen
  \bibfield  {author} {\bibinfo {author} {\bibfnamefont {A.}~\bibnamefont
  {Kay}}\ and\ \bibinfo {author} {\bibfnamefont {P.~J.}\ \bibnamefont
  {{Pemberton-Ross}}},\ }\href {\doibase 10.1103/PhysRevA.81.010301} {\bibfield
   {journal} {\bibinfo  {journal} {Physical Review A}\ }\textbf {\bibinfo
  {volume} {81}},\ \bibinfo {pages} {010301} (\bibinfo {year}
  {2010})}\BibitemShut {NoStop}%
\bibitem [{\citenamefont {Haselgrove}(2005)}]{haselgrove2005}%
  \BibitemOpen
  \bibfield  {author} {\bibinfo {author} {\bibfnamefont {H.~L.}\ \bibnamefont
  {Haselgrove}},\ }\href@noop {} {\bibfield  {journal} {\bibinfo  {journal}
  {Phys. Rev. A}\ }\textbf {\bibinfo {volume} {72}},\ \bibinfo {pages} {062326}
  (\bibinfo {year} {2005})}\BibitemShut {NoStop}%
\bibitem [{\citenamefont {Burgarth}\ and\ \citenamefont
  {Bose}(2005)}]{burgarth2005}%
  \BibitemOpen
  \bibfield  {author} {\bibinfo {author} {\bibfnamefont {D.}~\bibnamefont
  {Burgarth}}\ and\ \bibinfo {author} {\bibfnamefont {S.}~\bibnamefont
  {Bose}},\ }\href {\doibase 10.1088/1367-2630/7/1/135} {\bibfield  {journal}
  {\bibinfo  {journal} {New Journal of Physics}\ }\textbf {\bibinfo {volume}
  {7}},\ \bibinfo {pages} {135} (\bibinfo {year} {2005})}\BibitemShut {NoStop}%
\bibitem [{\citenamefont {Marletto}\ \emph {et~al.}(2012)\citenamefont
  {Marletto}, \citenamefont {Kay},\ and\ \citenamefont {Ekert}}]{marletto2012}%
  \BibitemOpen
  \bibfield  {author} {\bibinfo {author} {\bibfnamefont {C.}~\bibnamefont
  {Marletto}}, \bibinfo {author} {\bibfnamefont {A.}~\bibnamefont {Kay}}, \
  and\ \bibinfo {author} {\bibfnamefont {A.}~\bibnamefont {Ekert}},\ }\href
  {http://arxiv.org/abs/1202.2978} {\bibfield  {journal} {\bibinfo  {journal}
  {Quantum Inf. Comput.}\ }\textbf {\bibinfo {volume} {12}},\ \bibinfo {pages}
  {648} (\bibinfo {year} {2012})}\BibitemShut {NoStop}%
\bibitem [{\citenamefont {Calderbank}\ and\ \citenamefont
  {Shor}(1996)}]{calderbank1996}%
  \BibitemOpen
  \bibfield  {author} {\bibinfo {author} {\bibfnamefont {A.~R.}\ \bibnamefont
  {Calderbank}}\ and\ \bibinfo {author} {\bibfnamefont {P.~W.}\ \bibnamefont
  {Shor}},\ }\href {\doibase 10.1103/PhysRevA.54.1098} {\bibfield  {journal}
  {\bibinfo  {journal} {Physical Review A}\ }\textbf {\bibinfo {volume} {54}},\
  \bibinfo {pages} {1098} (\bibinfo {year} {1996})}\BibitemShut {NoStop}%
\bibitem [{\citenamefont {Steane}(1996)}]{steane1996}%
  \BibitemOpen
  \bibfield  {author} {\bibinfo {author} {\bibfnamefont {A.}~\bibnamefont
  {Steane}},\ }\href {\doibase 10.1098/rspa.1996.0136} {\bibfield  {journal}
  {\bibinfo  {journal} {Proceedings of the Royal Society of London A:
  Mathematical, Physical and Engineering Sciences}\ }\textbf {\bibinfo {volume}
  {452}},\ \bibinfo {pages} {2551} (\bibinfo {year} {1996})}\BibitemShut
  {NoStop}%
\bibitem [{\citenamefont {Blais}\ \emph {et~al.}(2004)\citenamefont {Blais},
  \citenamefont {Huang}, \citenamefont {Wallraff}, \citenamefont {Girvin},\
  and\ \citenamefont {Schoelkopf}}]{Blais2004}%
  \BibitemOpen
  \bibfield  {author} {\bibinfo {author} {\bibfnamefont {A.}~\bibnamefont
  {Blais}}, \bibinfo {author} {\bibfnamefont {R.}~\bibnamefont {Huang}},
  \bibinfo {author} {\bibfnamefont {A.}~\bibnamefont {Wallraff}}, \bibinfo
  {author} {\bibfnamefont {S.~M.}\ \bibnamefont {Girvin}}, \ and\ \bibinfo
  {author} {\bibfnamefont {R.~J.}\ \bibnamefont {Schoelkopf}},\ }\href
  {\doibase 10.1103/PhysRevA.69.062320} {\bibfield  {journal} {\bibinfo
  {journal} {Physical Review A}\ }\textbf {\bibinfo {volume} {69}},\ \bibinfo
  {pages} {062320} (\bibinfo {year} {2004})}\BibitemShut {NoStop}%
\bibitem [{\citenamefont {Majer}\ \emph {et~al.}(2007)\citenamefont {Majer},
  \citenamefont {Chow}, \citenamefont {Gambetta}, \citenamefont {Koch},
  \citenamefont {Johnson}, \citenamefont {Schreier}, \citenamefont {Frunzio},
  \citenamefont {Schuster}, \citenamefont {Houck}, \citenamefont {Wallraff},
  \citenamefont {Blais}, \citenamefont {Devoret}, \citenamefont {Girvin},\ and\
  \citenamefont {Schoelkopf}}]{majer2007}%
  \BibitemOpen
  \bibfield  {author} {\bibinfo {author} {\bibfnamefont {J.}~\bibnamefont
  {Majer}}, \bibinfo {author} {\bibfnamefont {J.~M.}\ \bibnamefont {Chow}},
  \bibinfo {author} {\bibfnamefont {J.~M.}\ \bibnamefont {Gambetta}}, \bibinfo
  {author} {\bibfnamefont {J.}~\bibnamefont {Koch}}, \bibinfo {author}
  {\bibfnamefont {B.~R.}\ \bibnamefont {Johnson}}, \bibinfo {author}
  {\bibfnamefont {J.~A.}\ \bibnamefont {Schreier}}, \bibinfo {author}
  {\bibfnamefont {L.}~\bibnamefont {Frunzio}}, \bibinfo {author} {\bibfnamefont
  {D.~I.}\ \bibnamefont {Schuster}}, \bibinfo {author} {\bibfnamefont {A.~A.}\
  \bibnamefont {Houck}}, \bibinfo {author} {\bibfnamefont {A.}~\bibnamefont
  {Wallraff}}, \bibinfo {author} {\bibfnamefont {A.}~\bibnamefont {Blais}},
  \bibinfo {author} {\bibfnamefont {M.~H.}\ \bibnamefont {Devoret}}, \bibinfo
  {author} {\bibfnamefont {S.~M.}\ \bibnamefont {Girvin}}, \ and\ \bibinfo
  {author} {\bibfnamefont {R.~J.}\ \bibnamefont {Schoelkopf}},\ }\href
  {\doibase 10.1038/nature06184} {\bibfield  {journal} {\bibinfo  {journal}
  {Nature}\ }\textbf {\bibinfo {volume} {449}},\ \bibinfo {pages} {443}
  (\bibinfo {year} {2007})}\BibitemShut {NoStop}%
\bibitem [{\citenamefont {Bialczak}\ \emph {et~al.}(2010)\citenamefont
  {Bialczak}, \citenamefont {Ansmann}, \citenamefont {Hofheinz}, \citenamefont
  {Lucero}, \citenamefont {Neeley}, \citenamefont
  {{O{\textquoteright}Connell}}, \citenamefont {Sank}, \citenamefont {Wang},
  \citenamefont {Wenner}, \citenamefont {Steffen}, \citenamefont {Cleland},\
  and\ \citenamefont {Martinis}}]{Bialczak2010}%
  \BibitemOpen
  \bibfield  {author} {\bibinfo {author} {\bibfnamefont {R.~C.}\ \bibnamefont
  {Bialczak}}, \bibinfo {author} {\bibfnamefont {M.}~\bibnamefont {Ansmann}},
  \bibinfo {author} {\bibfnamefont {M.}~\bibnamefont {Hofheinz}}, \bibinfo
  {author} {\bibfnamefont {E.}~\bibnamefont {Lucero}}, \bibinfo {author}
  {\bibfnamefont {M.}~\bibnamefont {Neeley}}, \bibinfo {author} {\bibfnamefont
  {A.~D.}\ \bibnamefont {{O{\textquoteright}Connell}}}, \bibinfo {author}
  {\bibfnamefont {D.}~\bibnamefont {Sank}}, \bibinfo {author} {\bibfnamefont
  {H.}~\bibnamefont {Wang}}, \bibinfo {author} {\bibfnamefont {J.}~\bibnamefont
  {Wenner}}, \bibinfo {author} {\bibfnamefont {M.}~\bibnamefont {Steffen}},
  \bibinfo {author} {\bibfnamefont {A.~N.}\ \bibnamefont {Cleland}}, \ and\
  \bibinfo {author} {\bibfnamefont {J.~M.}\ \bibnamefont {Martinis}},\ }\href
  {\doibase 10.1038/nphys1639} {\bibfield  {journal} {\bibinfo  {journal}
  {Nature Physics}\ }\textbf {\bibinfo {volume} {6}},\ \bibinfo {pages} {409}
  (\bibinfo {year} {2010})}\BibitemShut {NoStop}%
\bibitem [{\citenamefont {Plantenberg}\ \emph {et~al.}(2007)\citenamefont
  {Plantenberg}, \citenamefont {de~Groot}, \citenamefont {Harmans},\ and\
  \citenamefont {Mooij}}]{Plantenberg2007}%
  \BibitemOpen
  \bibfield  {author} {\bibinfo {author} {\bibfnamefont {J.~H.}\ \bibnamefont
  {Plantenberg}}, \bibinfo {author} {\bibfnamefont {P.~C.}\ \bibnamefont
  {de~Groot}}, \bibinfo {author} {\bibfnamefont {C.~J. P.~M.}\ \bibnamefont
  {Harmans}}, \ and\ \bibinfo {author} {\bibfnamefont {J.~E.}\ \bibnamefont
  {Mooij}},\ }\href {\doibase 10.1038/nature05896} {\bibfield  {journal}
  {\bibinfo  {journal} {Nature}\ }\textbf {\bibinfo {volume} {447}},\ \bibinfo
  {pages} {836} (\bibinfo {year} {2007})}\BibitemShut {NoStop}%
\bibitem [{\citenamefont {{Garc\'{i}a-Ripoll}}\ and\ \citenamefont
  {Cirac}(2003)}]{Garcia-Ripoll2003}%
  \BibitemOpen
  \bibfield  {author} {\bibinfo {author} {\bibfnamefont {J.~J.}\ \bibnamefont
  {{Garc\'{i}a-Ripoll}}}\ and\ \bibinfo {author} {\bibfnamefont {J.~I.}\
  \bibnamefont {Cirac}},\ }\href {\doibase 10.1088/1367-2630/5/1/376}
  {\bibfield  {journal} {\bibinfo  {journal} {New Journal of Physics}\ }\textbf
  {\bibinfo {volume} {5}},\ \bibinfo {pages} {76} (\bibinfo {year}
  {2003})}\BibitemShut {NoStop}%
\bibitem [{\citenamefont {Karbach}\ and\ \citenamefont
  {Stolze}(2005)}]{karbach2005}%
  \BibitemOpen
  \bibfield  {author} {\bibinfo {author} {\bibfnamefont {P.}~\bibnamefont
  {Karbach}}\ and\ \bibinfo {author} {\bibfnamefont {J.}~\bibnamefont
  {Stolze}},\ }\href {\doibase 10.1103/PhysRevA.72.030301} {\bibfield
  {journal} {\bibinfo  {journal} {Physical Review A}\ }\textbf {\bibinfo
  {volume} {72}},\ \bibinfo {pages} {030301} (\bibinfo {year}
  {2005})}\BibitemShut {NoStop}%
\bibitem [{\citenamefont {Di~Franco}\ \emph {et~al.}(2008)\citenamefont
  {Di~Franco}, \citenamefont {Paternostro},\ and\ \citenamefont
  {Kim}}]{difranco2008}%
  \BibitemOpen
  \bibfield  {author} {\bibinfo {author} {\bibfnamefont {C.}~\bibnamefont
  {Di~Franco}}, \bibinfo {author} {\bibfnamefont {M.}~\bibnamefont
  {Paternostro}}, \ and\ \bibinfo {author} {\bibfnamefont {M.~S.}\ \bibnamefont
  {Kim}},\ }\href {\doibase 10.1103/PhysRevLett.101.230502} {\bibfield
  {journal} {\bibinfo  {journal} {Physical Review Letters}\ }\textbf {\bibinfo
  {volume} {101}},\ \bibinfo {pages} {230502} (\bibinfo {year}
  {2008})}\BibitemShut {NoStop}%
\bibitem [{\citenamefont {Albanese}\ \emph {et~al.}(2004)\citenamefont
  {Albanese}, \citenamefont {Christandl}, \citenamefont {Datta},\ and\
  \citenamefont {Ekert}}]{albanese2004}%
  \BibitemOpen
  \bibfield  {author} {\bibinfo {author} {\bibfnamefont {C.}~\bibnamefont
  {Albanese}}, \bibinfo {author} {\bibfnamefont {M.}~\bibnamefont
  {Christandl}}, \bibinfo {author} {\bibfnamefont {N.}~\bibnamefont {Datta}}, \
  and\ \bibinfo {author} {\bibfnamefont {A.}~\bibnamefont {Ekert}},\ }\href
  {\doibase 10.1103/PhysRevLett.93.230502} {\bibfield  {journal} {\bibinfo
  {journal} {Physical Review Letters}\ }\textbf {\bibinfo {volume} {93}},\
  \bibinfo {pages} {230502} (\bibinfo {year} {2004})}\BibitemShut {NoStop}%
\bibitem [{\citenamefont {Kay}(2007)}]{kay2007}%
  \BibitemOpen
  \bibfield  {author} {\bibinfo {author} {\bibfnamefont {A.}~\bibnamefont
  {Kay}},\ }\href {\doibase 10.1103/PhysRevLett.98.010501} {\bibfield
  {journal} {\bibinfo  {journal} {Physical Review Letters}\ }\textbf {\bibinfo
  {volume} {98}},\ \bibinfo {pages} {010501} (\bibinfo {year}
  {2007})}\BibitemShut {NoStop}%
\bibitem [{Note1()}]{Note1}%
  \BibitemOpen
  \bibinfo {note} {Alternatively, one could simply update the stabilizers of
  the code.}\BibitemShut {Stop}%
\bibitem [{\citenamefont {Aliferis}\ \emph {et~al.}(2006)\citenamefont
  {Aliferis}, \citenamefont {Gottesman},\ and\ \citenamefont
  {Preskill}}]{aliferis2006}%
  \BibitemOpen
  \bibfield  {author} {\bibinfo {author} {\bibfnamefont {P.}~\bibnamefont
  {Aliferis}}, \bibinfo {author} {\bibfnamefont {D.}~\bibnamefont {Gottesman}},
  \ and\ \bibinfo {author} {\bibfnamefont {J.}~\bibnamefont {Preskill}},\
  }\href {http://arxiv.org/abs/quant-ph/0504218} {\bibfield  {journal}
  {\bibinfo  {journal} {Quant. Inf. Comput.}\ }\textbf {\bibinfo {volume}
  {6}},\ \bibinfo {pages} {97} (\bibinfo {year} {2006})},\ \bibinfo {note}
  {quant. Inf. Comput. 6 (2006) 97-165}\BibitemShut {NoStop}%
\bibitem [{\citenamefont {Burrell}\ \emph {et~al.}(2009)\citenamefont
  {Burrell}, \citenamefont {Eisert},\ and\ \citenamefont
  {Osborne}}]{burrell2009}%
  \BibitemOpen
  \bibfield  {author} {\bibinfo {author} {\bibfnamefont {C.~K.}\ \bibnamefont
  {Burrell}}, \bibinfo {author} {\bibfnamefont {J.}~\bibnamefont {Eisert}}, \
  and\ \bibinfo {author} {\bibfnamefont {T.~J.}\ \bibnamefont {Osborne}},\
  }\href {\doibase 10.1103/PhysRevA.80.052319} {\bibfield  {journal} {\bibinfo
  {journal} {Physical Review A}\ }\textbf {\bibinfo {volume} {80}},\ \bibinfo
  {pages} {052319} (\bibinfo {year} {2009})}\BibitemShut {NoStop}%
\bibitem [{Note2()}]{Note2}%
  \BibitemOpen
  \bibinfo {note} {P.\ M\'ajek, Diploma Thesis, Comenius University
  (2005)}\BibitemShut {NoStop}%
\bibitem [{\citenamefont {Ioffe}\ and\ \citenamefont
  {Mezard}(2007)}]{ioffe2007}%
  \BibitemOpen
  \bibfield  {author} {\bibinfo {author} {\bibfnamefont {L.}~\bibnamefont
  {Ioffe}}\ and\ \bibinfo {author} {\bibfnamefont {M.}~\bibnamefont {Mezard}},\
  }\href {\doibase 10.1103/PhysRevA.75.032345} {\bibfield  {journal} {\bibinfo
  {journal} {Physical Review A}\ }\textbf {\bibinfo {volume} {75}},\ \bibinfo
  {pages} {032345} (\bibinfo {year} {2007})}\BibitemShut {NoStop}%
\bibitem [{Note3()}]{Note3}%
  \BibitemOpen
  \bibinfo {note} {A.\ Kay (2016): full\protect \_test.nb. figshare.
  doi:10.6084/m9.figshare.2070181}\BibitemShut {NoStop}%
\bibitem [{\citenamefont {Yung}(2006)}]{yung2006}%
  \BibitemOpen
  \bibfield  {author} {\bibinfo {author} {\bibfnamefont {M.}~\bibnamefont
  {Yung}},\ }\href {\doibase 10.1103/PhysRevA.74.030303} {\bibfield  {journal}
  {\bibinfo  {journal} {Physical Review A}\ }\textbf {\bibinfo {volume} {74}},\
  \bibinfo {pages} {030303} (\bibinfo {year} {2006})}\BibitemShut {NoStop}%
\bibitem [{\citenamefont {Kay}(2006)}]{kay2006}%
  \BibitemOpen
  \bibfield  {author} {\bibinfo {author} {\bibfnamefont {A.}~\bibnamefont
  {Kay}},\ }\href {\doibase 10.1103/PhysRevA.73.032306} {\bibfield  {journal}
  {\bibinfo  {journal} {Physical Review A}\ }\textbf {\bibinfo {volume} {73}},\
  \bibinfo {pages} {032306} (\bibinfo {year} {2006})}\BibitemShut {NoStop}%
\bibitem [{\citenamefont {Kitaev}(2003)}]{kitaev2003}%
  \BibitemOpen
  \bibfield  {author} {\bibinfo {author} {\bibfnamefont {A.}~\bibnamefont
  {Kitaev}},\ }\href {\doibase 10.1016/S0003-4916(02)00018-0} {\bibfield
  {journal} {\bibinfo  {journal} {Annals of Physics}\ }\textbf {\bibinfo
  {volume} {303}},\ \bibinfo {pages} {2} (\bibinfo {year} {2003})}\BibitemShut
  {NoStop}%
\bibitem [{\citenamefont {Dennis}\ \emph {et~al.}(2002)\citenamefont {Dennis},
  \citenamefont {Kitaev}, \citenamefont {Landahl},\ and\ \citenamefont
  {Preskill}}]{dennis2002}%
  \BibitemOpen
  \bibfield  {author} {\bibinfo {author} {\bibfnamefont {E.}~\bibnamefont
  {Dennis}}, \bibinfo {author} {\bibfnamefont {A.}~\bibnamefont {Kitaev}},
  \bibinfo {author} {\bibfnamefont {A.}~\bibnamefont {Landahl}}, \ and\
  \bibinfo {author} {\bibfnamefont {J.}~\bibnamefont {Preskill}},\ }\href
  {\doibase 10.1063/1.1499754} {\bibfield  {journal} {\bibinfo  {journal}
  {Journal of Mathematical Physics}\ }\textbf {\bibinfo {volume} {43}},\
  \bibinfo {pages} {4452} (\bibinfo {year} {2002})}\BibitemShut {NoStop}%
\bibitem [{\citenamefont {Kay}(2014)}]{kay2014}%
  \BibitemOpen
  \bibfield  {author} {\bibinfo {author} {\bibfnamefont {A.}~\bibnamefont
  {Kay}},\ }\href {\doibase 10.1103/PhysRevA.89.032328} {\bibfield  {journal}
  {\bibinfo  {journal} {Physical Review A}\ }\textbf {\bibinfo {volume} {89}},\
  \bibinfo {pages} {032328} (\bibinfo {year} {2014})}\BibitemShut {NoStop}%
\end{thebibliography}%

\end{document}